\input epsf
\documentstyle[aps,eqsecnum]{revtex}
\preprint{\vbox{\baselineskip=12pt
\rightline{CGPG-96/11-2}
\rightline{gr-qc/9611034}}}

\def\be{\nopagebreak[3]\begin{equation}}
\def\ee{\end{equation}}
\def\ba{\nopagebreak[3]\begin{eqnarray}}
\def\ea{\end{eqnarray}}
\def\nl{\nonumber \\}
\def\ni{\noindent}
\def\a{\alpha}
\def\b{\beta}
\def\c{\gamma}
\def\d{\delta}

\def\ve{\varepsilon}
\def\f{\phi}
\def\g{\raisebox{.4ex}{$\gamma$}}

\def\m{\mu}
\def\n{\nu}

\newcommand{\teta}{\rlap{\lower2ex\hbox{$\,\tilde{}$}}\eta{}}
\newcommand{\tiN}{\rlap{\lower2ex\hbox{$\,\tilde{}$}}N{}}

\newcommand{\tE}{\tilde{E}}
\newcommand{\hU}{\hat{U}_{\varphi}}
\newcommand{\amodg}{\overline{\cal A/G}}
\begin{document}
\draft
\title{On Diffeomorphism Invariance\\
 for Lattice Theories}
\author {Alejandro Corichi
\thanks{Electronic address: corichi@phys.psu.edu}
and Jos\'e A. Zapata
\thanks{Electronic address: zapata@phys.psu.edu}
}
\address{Center for Gravitational Physics and Geometry \\
Department of Physics, The Pennsylvania State University \\
University Park, PA 16802, USA}
\maketitle
\begin{abstract}

We consider the role of the diffeomorphism 
constraint in the quantization of lattice formulations
of diffeomorphism invariant theories of connections. 
It has been argued that in working with abstract lattices,
one  automatically takes care of the diffeomorphism constraint in the 
quantum theory. We use two systems in order to show that 
imposing the diffeomorphism constraint is imperative to 
obtain a physically acceptable quantum theory.
First, we consider $2+1$ gravity  where
an exact lattice formulation is available.
Next,  general theories of connections for compact 
gauge groups are treated,
where the quantum theories are known --for both the 
continuum and the lattice-- and can be compared.  
 
\end{abstract}
\pacs{PACS numbers: 0460Ds, 0460Kz. 0460Nc}

\section{Introduction}

Within the canonical quantization of diffeomorphism invariant theories
of connections, at some stage one is forced to deal with the
diffeomorphism constraint. Examples of such theories are given by
gravity in 3 and 4 dimensions, when formulated as theories of
connections \cite{ab87,witt1}.  As first noticed by Rovelli and Smolin
\cite{carlo-leeDIFF}, when the basic observables are the (traces of)
 holonomies of the connection $W_{\g}={\rm tr}(h_{\g})$ along 
loops $\g$ in the manifold, finite diffeomorphisms act simply by
shifting the loops as prescribed by the mapping $\f$: $U_{\f} \circ
W_{\g}=W_{\f(\g)}$.  Further developments of this idea rigorously
solve the diffeomorphism constraint in the continuum. For a complete
discussion of the quantization of diffeomorphism invariant theories
see  \cite{alm2t}.

In theories that have discretized space from the outset, the issue of
the diffeomorphism constraint has a special character. In some 
of these discretized theories, space is
replaced with an abstract lattice; that is, a lattice that does not
``reside'' in a manifold \cite{wael1,renata}. Hence the diffeomorphism 
symmetry is lost in this approach.  However, an abstract lattice 
$L$ can be viewed as
an equivalence class of embedded lattices $[L]$ under the equivalence
relation defined by the diffeomorphism group, in the sense used in
knot theory to define knot classes. It has been argued that this
picture implies that the theory defined on an abstract lattice is
manifestly diffeomorphism invariant, and therefore, there is no need
to further impose the diffeomorphism constraint \cite{renata}.  A
different approach considers that even after discretization, the phase
space associated to the abstract lattice still has a reminiscent
symmetry related to the diffeomorphism constraint. This strategy has
been successful in treating theories without local degrees of freedom
like $2+1$ gravity and $B\wedge F$ theories \cite{wael1,h-j}.  A third
strategy on the lattice, embedded in a manifold,  is to require
invariance under a discrete symmetry such that diffeomorphism
invariance is recovered in the continuum limit \cite{rent-smo,jorge-rodolfo}.
We will focus our attention on the first two strategies for discrete
theories.

The physically meaningful aspects of a lattice theory are the ones
regarding its continuum limit. This makes comparison between different
lattice approaches a delicate issue.  To illustrate the discrepancy
between the two lattice approaches described above, we shall consider
$2+1$ dimensional gravity.  In this framework the role of the continuum 
limit is not very significant because it is a theory with only global
degrees of freedom that are fully contained in the lattice theory
already.  We show that by assuming that a theory based on abstract
lattices is manifestly diffeomorphism invariant one fails to
`subtract' the appropriate degrees of freedom from the unconstrained
theory. Therefore, the resulting theory is not the one we wanted to
quantize; it has too many degrees of freedom.

In the second part of the article, we consider diffeomorphism invariant
theories for compact gauge groups. For this class of theories, there 
is a  well defined formalism to treat the quantization both in
the continuum and in the lattice. For completeness, we first review 
this formalism and then show that, independently of the details of
the continuum limit one might choose to take, there is still a residual
symmetry associated to diffeomorphisms.

The article is organized as follows. Section II reviews the classical
description of $2+1$ gravity as a theory of connections, both in
the continuum and in the  lattice formulation. Section III is devoted to
discussing the quantum theory for general theories of connections with
compact gauge groups. We review first the quantization in the continuum
and then in the lattice. It is shown that
one must impose all constraints, including the ones associated to
diffeomorphism, in order to recover the physical theory. We conclude
in Sec. IV with a discussion.

\section{$2+1$ gravity in the connection-dynamics formulation}

\subsection{$2+1$ in the continuum}

In this part we give a brief review of the continuum description
 following the notation of \cite{ash-rom}. For
a detailed exposition we refer the reader to the original papers
\cite{witt1,ash-rom}.  
 Let $M$ be a 3-dimensional manifold of the
form $\Sigma\times R$, where $\Sigma$ is a compact, oriented
2-manifold. In the Hamiltonian formulation of the theory, the phase
space $\Gamma$ can be coordinatized by pairs $(A_a^I,\tE^b_J)$ on
$\Sigma$. Here, $A_a^I$ is the pull-back to $\Sigma$ of the
3-dimensional Lorentz connection. As a field on the hyper-surface it is
a one-form with values in the Lorentz Lie-algebra $so(2,1)$. The
canonical momenta is given by the vector densities $\tE^a_I$. The
1-forms $e_{aI} =\teta_{ab}{\tE}^b_I$ are the pull-back to
$\Sigma$ of the co-triad ${}^3\!e^I_a$ on $M$. (Throughout we will 
raise and lower the internal index $I$ with the Minkowski metric
$\eta_{IJ}={\rm diag}(-,+,+)$). 
The canonical pair of variables
satisfy the usual Poisson bracket relations,
\be
\left\{A_a^I(x),\tE^b_J(y)\right\}=\delta^I_J\delta^b_a\delta^3(x,y).
\ee
From this variables we can recover the 2-dimensional metric $g_{ab}$ on
$\Sigma$ of the usual geometrodynamical formulation as follows:
\be
g_{ab}=\eta_{IJ}e_a^Ie_b^J.
\ee
Since we want to consider positive definite metrics $g_{ab}$, we
should impose the restriction on  the phase space coordinates that
$\tE$ be non-degenerate.

There are two constraints on the phase space, given by
\be
{\cal D}_a \tE^a_I=0\;\;\;\;\;\;{\rm and}
\;\;\;\;\;\;F^I_{ab}=0,
\ee
where ${\cal D}$ denotes the generalized covariant derivative defined by
$A_a^I$, namely ${\cal D}_av_{bJ}:=\partial_av_{bJ}+{\Gamma_{ab}}^cv_{cJ}+
{\varepsilon_{JI}}^KA^I_av_{cK}$.
$F^I_{ab}$ is the curvature of the connection: $F^I_{ab}=2\partial_{[a}
A_{b]}^I+[A_a,A_b]^I$. The first constraint is known as the Gauss law.
In order to analyze what the constraints generate, let us define the
smeared functions on  $\Gamma$,
\ba
G[v]&:=&\int_\Sigma v^I{\cal D}_a \tE^a_I,\nl
F[\alpha]&:=&\int_\Sigma \tilde{\alpha}_I^{ab}F^I_{ab}.\label{wittcon}
\ea  
The infinitesimal canonical transformation generated by the constraint
functions will depend, of course, on the smearing fields $(v,\alpha)$.

The Gauss constraint $G[v]$ generates gauge transformations on
the connection $A_a^I$ and `rotations' in the momenta $\tE^a_I$. The
constraint function $F[\alpha]$ leaves the connection invariant and `shifts'
the conjugate momenta. We know that this constrained system
is equivalent to the geometrodynamical description \cite{Berg}. Therefore, an
appropriate combination of the constraint functions should generate the
induced action of diffeomorphisms on $\Sigma$. Let us start by considering
the connection $A$. Since the curvature constraint does not depend on the
momenta $E$, it leaves the connection unchanged. That means that the
only way one can generate diffeomorphism 
on the connection is via the Gauss constraint, with
an appropriate smearing field $v^I$.

Consider the vector field $V^a$ on $\Sigma$,
as generator of infinitesimal diffeomorphisms. The function,
\be
\int_\Sigma (A_b^I V^b){\cal D}_a\tE^a_I,
\ee
generates the required action on $A$ since,
\ba
\left\{A_a^I,\int_\Sigma (A_b^J V^b){\cal D}_c\tE^c_J\right\}&=&
-{\cal D}_a(A_b^JV^b),\nl
&=&-V^bF^I_{ab}-{\cal L}_V A^a_I\approx-{\cal L}_V A^a_I.
\ea
More generally, the generator of spatial diffeomorphisms is given by,
\be
D[V]=\int_\Sigma V^b(A_b^I{\cal D}_a\tE^a_I-\tE^a_IF^I_{ab}). 
\label{d[v]}
\ee
Note that the second term, where we have chosen
$\tilde{\alpha}^{ab}_I=V^{[b}\tE^{a]}_I$, can also be interpreted as
the projection of the curvature constraint along the vector density 
$\tE^a_I$.
One way to impose the curvature constraint is to require that the
projections of the curvature in every internal direction vanish. In
fact, after imposing (\ref{d[v]}) we need only one more projection of
the curvature constraint.
\be
H[N]:=\int_\Sigma\tiN{\varepsilon_I}^{JK}\tE^a_J\tE^b_K\,F^I_{ab},
\ee
where $\tiN$ is a scalar density of weight $-1$. More precisely, 
provided that the co-triad $e_a^I$ is non-degenerate, the
Witten constraints (\ref{wittcon}) and the Ashtekar
constraints (below) are equivalent \cite{Berg}.  
\ba
{\cal D}_a\tE^a_I&=&0,\nl
\tE^b_IF^I_{ab}&=&0,\label{ashtcon}\\
{\varepsilon_I}^{JK}\tE^a_J\tE^b_K\,F^I_{ab}&=&0.\nonumber
\ea
The Witten constraints can be seen as `covariant' since there is no
explicit decomposition of the 3-dimensional diffeomorphisms into a
`spatial' and `orthogonal' components. 
The Ashtekar constraints on the
other hand, do exhibit this decomposition. They are more closely
related to the constraints one encounters in the 3+1-dimensional
theory.  When one allows for degenerate
triads (and therefore singular induced metrics), the constraints
(\ref{ashtcon}) do not imply the vanishing of the curvature
$F^I_{ab}$. The structure of such extended phase space has been
studied in \cite{ferymad}.

\subsection{2+1 gravity in the Lattice Formulation}

In this part, we recall the lattice formulation of 
$2+1$ gravity due to Waelbroeck\cite{wael1}. 
For the convenience of the reader the roles of the 
dual lattice and the lattice itself are reversed,
with respect to \cite{wael1}. 
The starting point is a two dimensional abstract lattice of arbitrary 
valence; the $N_0$ vertices of the lattice are labeled by greek indices 
$(\a) ,(\b), \ldots$, the $N_1$ links are labeled as pairs of vertices 
$(\a \b), (\b \g), \ldots$ and the $N_2$ faces are labeled as 
$f_1, f_2, \ldots$. To each vertex $(\a)$ of the lattice
we assign a three dimensional Minkowski frame (a copy of the 
Lie algebra $so(2,1)$); 
the role of the connection is played by 3-dimensional Lorentz matrices
${\bf M}_{(\a \b)}$ that define parallel transport from vertex $(\b)$ 
to vertex $(\a )$. In this way the configuration space of the lattice 
theory is $SO(2,1)^{N_1}$. 
The momenta ${\bf E}_{(\a \b)}$, represented by vectors of the 
Lie algebra, label the left and  right-invariant 
vector fields of $SO(2,1)$ as can be seen in the Poisson 
algebra 
\ba
\{ {E_{(\a \b)}}^I,{E_{(\a \b)}}^J \} &=& 
{\ve ^{IJ}}_K{E_{(\a \b)}}^K, \\
\{ {E_{(\a \b)}}^I , {{M_{(\a \b)}}^J}_K \} &=& 
{\ve ^{IJ}}_L{{M_{(\a \b)}}^L}_K, \\
\{ {E_{(\a \b)}}^I,{{M_{(\b \a)}}^J}_K \} &=&
{{\ve ^I}_K}^L {{M_{(\b \a)}}^J}_L. 
\ea
The rest of the non-vanishing brackets follow from the 
identities 
\ba
{{M_{(\a \b)}}^I}_J{{M_{(\b \a)}}^J}_K&=&\delta^I_K,\\
{{M_{(\a \b)}}^I}_J{M_{(\a \b)}}^{JK}&=&\eta^{IK},\\ 
{{M_{(\a \b)}}^I}_J{E_{(\b \a)}}^J&=&-{E_{(\a \b)}}^I. \label{e=-me}
\ea
Identity (\ref{e=-me}) plays a very important role; first, 
it relates the Lie algebra vectors that label the 
left and right-invariant vector fields to give three independent 
momentum coordinates per link of the lattice. 
Second, thanks to (\ref{e=-me}) a geometric interpretation of 
the variables is possible. 
The links of the dual lattice are placed in the Minkowski frames of 
the vertices of the original lattice (dual faces) according to the 
variables ${\bf E}_{(\a \b)}$. And after Gauss's law is satisfied 
\be
{J(\a)}^I:={E_{(\a \b)}}^I+ {E_{(\a \c)}}^I+\ldots= 0, 
\label{j=0}
\ee
the faces of the dual lattice close. To make this lattice gauge 
theory describe vacuum $2+1$ gravity we have to require that the curvature 
vanishes; in other words, the parallel transport around every face of the 
lattice (around every vertex of the dual lattice) must be the identity. 
That is, 
\be
{{W(f_1)}^I}_J:={{({\bf M}_{\a \b}{\bf M}_{\b \m}\ldots
{\bf M}_{\n \a})}^I}_J={{\d}^I}_J, \label{p=0}
\ee
where the boundary of face $f_1$ is composed by the links 
$(\a \b),(\b \m),\ldots,(\n \a)$. Therefore the holonomy around 
{\em every contractible loop} of the lattice is restricted to be 
the identity. The curvature constraint can 
be replaced by $P(f_1)^I:=\frac{1}{2}
{{\ve ^I}_J}^K{{W(f_1)}^J}_K=0$. The constraints are 
first class, and generate Lorentz transformations at the frame $(\a)$ and
translations of vertex $(f_1)$ of the dual lattice respectively \cite{wael1}.
\ba
\left\{{J(\a)}^I,{E_{(\a \b)}}^J\right\}&=&{{\ve }^{IJ}}_K
{E_{(\a \b)}}^K,\\
\left\{{J(\a)}^I,{{M_{(\a \b)}}^J}_K\right\}&=&{{\ve }^{IJ}}_L
{{M_{(\a \b)}}^J}_L,\\
\left\{\xi^IP(f_1)_I,{E_{(\a \b)}}^J\right\}&\approx&\xi^J.
\ea
(The weak equality indicates that we are restricting ourselves to the
constraint surface).
Waelbroeck's constraints (\ref{j=0}), (\ref{p=0}) 
are the precise analog of the Witten constraints
(\ref{wittcon}) in the continuum. 

When the $E$'s are non-degenerate the constraints can be written in an 
equivalent Ashtekar form. More precisely, when at every vertex
$(\a)$  the vector space generated by 
${\bf E}_{(\a \b)}, {\bf E}_{(\a \c)}, \ldots $ 
is at least two dimensional, the projected curvature constraints 
\ba 
H_{(\a \b)} &:=&B(\a)^I {E_{(\a \b)}}_I =0, \label{ha}\\
H(\a)&:=& \ve ^{IJK} B(\a)_I ({E_{(\a \b)}}_J 
{E_{(\a \c)}}_K +{E_{(\a \c)}}_J {E_{(\a \d)}}_K+ 
\ldots {E_{(\a \n)}}_J {E_{(\a \b)}}_K )=0, \label{h}
\ea 
imply $B(\a)^I:=\sum_{\a \in fi} P(fi)^I=0$ %
\footnote{
the sum is over the curvature vectors of all the faces that contain 
vertex $(\a)$; the orientation of the faces is the one induced by 
the orientation of the lattice. 
}. 
Hence, the  constraints (\ref{j=0}), (\ref{ha}) and (\ref{h}), that 
are of the Ashtekar type, form a first-class system. 

The question of whether the projected constraints (\ref{ha}), (\ref{h}) 
are equivalent to the covariant constraint (\ref{p=0}) is studied in 
the appendix. We consider the case of a square $M \times N$ 
lattice with periodic boundary conditions, i.e. $T^2$ represented by 
rectangular grid; in this case we prove that the two sets of 
constraints are equivalent if and only if $M$ and $N$ are odd. 
Because of its interest in the case of a possible relation with 
Regge calculus, we consider the case of three-valent lattices. These 
are lattices whose duals are triangular lattices. 
Again there are cases of lattices where the covariant and projected 
versions of the constraints are not equivalent. However, given 
one of this lattices $L$ where the two sets of constraints are not 
equivalent we construct a new tri-valent lattice $L^\prime$ where 
the the projected constraints (\ref{ha}), (\ref{h}) are equivalent 
to the covariant constraints (\ref{p=0}). We know that any surface 
can be represented by a three-valent lattice (because any surface can 
be triangulated). Therefore we can represent any 
two dimensional space by a lattice where the covariant constraints 
are equivalent to the projected constraints, 
in the sector where the $E$'s are non-degenerate. 

The constraint (\ref{h}) is the lattice version of the Hamiltonian 
constraint (it is linear in the curvature and quadratic in the 
momenta), while (\ref{ha}) plays the role of the diffeomorphism 
constraint (it is linear in the curvature and in the momenta). 
We see that algebraically (\ref{ha}) 
takes care of the residual diffeomorphism symmetry. Geometrically 
the constraint is the generator of phase space symmetries that 
{\em do} correspond to translations of the vertices (dual faces) of 
the lattice. 

In $2+1$ gravity the lattice theory is exact in the sense that it contains 
all the true degrees of freedom of the continuum theory. 
The number of variables of the phase space 
is $6N_1$, while the number of first-class constraints is $3N_0 +3N_2$. 
Hence the dimension of the reduced phase space is 
\be 
6N_1 - 2( 3N_0 + 3N_2) = 12g -12, 
\ee 
where $g$ is the genus of space. In this lattice approach two 
different lattices with the same topology give rise to two 
descriptions of the same reduced phase space.

In this subsection we have recalled Waelbroeck's exact lattice formulation 
of $2+1$ gravity, and we have rewritten the constraints in a projected 
form. By the counting of degrees of freedom given above, we see that in a 
quantum theory 
based on an abstract lattice one needs to impose a diffeomorphism 
constraint to reproduce a classical theory with the 
correct number of degrees of freedom. This is the first remark of 
this paper.

\section{quantization}

In this section, we will obtain the second result of the article.
The section is divided in to three parts. In the first one, we give
some heuristic arguments that suggest the existence of a residual
diffeomorphism symmetry for theories defined on an abstract lattice. 
In the second part,
we make a small detour and recall, for the continuum, the quantization
of theories of connections that have a compact gauge group,
such as gravity in $3+1$ dimensions and Euclidean $2+1$ gravity.
In the last part, we consider the quantization of the 
lattice formulation and compare it to the continuum.

\subsection{Heuristics}

The issue of diffeomorphism invariance in the quantization of the 
lattice theory is a subtle one, since it is not apriory clear  how
to achieve a continuum limit which is compatible with diffeomorphism 
invariance. The proposal that by working on an abstract lattice one
automatically solves the diffeomorphism constraint is an  attractive one.
The idea is that by  quantizing the gauge theory on an abstract
lattice, one gets a family of states that 
should be mapped to some diffeomorphism invariant states after the
continuum limit is taken. 
 
In order to test the full physical viability of the theory one should
wait for the construction of a consistent continuum limit and then
compare its predictions with the `known physics'. However, there are
general arguments that one can give,  that do {\it  not} depend on the
details of the particular continuum limit that one selects. That is the
aim of the second half of the paper. We want to argue that,
roughly speaking,
there are states defined over the lattice that should be identified in
any continuum limit that recovers diffeomorphism invariance, and that
are clearly distinct from the view-point of the abstract lattice. 
Therefore, there is a residual diffeomorphism symmetry 
even when working in the abstract lattice. 

Let us consider, for instance, the lattice of Fig.1, together with two
loops $\gamma_1$ and $\gamma_2$. As loops defined on the lattice, they
are clearly different. However, when seen as  embedded in a manifold, 
there is a diffeomorphism that maps $\gamma_1$ to $\gamma_2$.
 One could expect that
in a limit that recovers spatial diffeomorphisms, states defined over
$\gamma_1$ and $\gamma_2$ are identified.

\hskip 1in
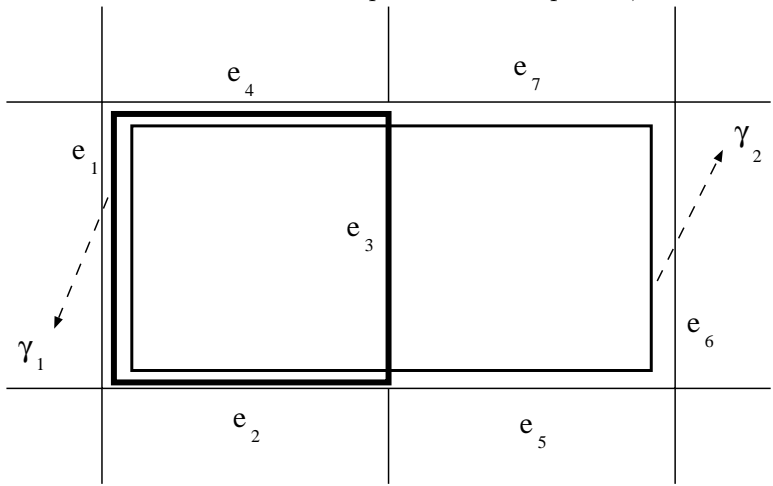
\epsfbox{corichi01.eps}
\bigskip

\noindent 
{\small {\bf Fig. 1}
The loops $\c_1 , \c_2$ are given by 
$\c_1 = e_1 \circ e_2 \circ e_3 \circ e_4$ and 
$\c_2 = e_1 \circ e_2 \circ e_5 \circ e_6 \circ e_7 \circ e_4$ where 
$e_1 , \ldots , e_7 \in L$. The loops are diffeomorphic but the states 
defined on them are different.
}

To make these intuitive arguments conclusive, we shall go in to
the details of the quantization on the continuum, and later compare it
to the quantization of the abstract lattice.

\subsection{Quantization in the Continuum}

In the quantization of the continuum theory, we can follow the procedure 
introduced in \cite{alm2t} for $3+1$ dimensions . The main
steps are the introduction of a kinematical Hilbert space where the
quantum constraints are to be defined, and the construction of the
Hilbert space of solutions to the Gauss and diffeomorphism constraint. 
One basic assumption in this program is that the holonomies of the 
connection $A_a^I$ are well defined operators in the kinematical Hilbert 
space. This assumption leads to $\overline{\cal A}$ that 
`completes' the space of smooth connections ${\cal A}$, to form 
the {\it quantum configuration space} of `generalized connections'.

Here we review the characterization of $\overline{\cal A}$ 
as a limit of a family of configuration
spaces living on finite `floating' lattices \cite{AL1,AL2}.
To each lattice $L$ {\it embedded} in $\Sigma$ 
we assign a configuration space ${\cal A}_L = G^{N_1}$. A point in 
${\cal A}_L$ is represented by  $(g_1,g_2,\ldots,g_{N_1})$, each group 
element $g_i$ is to be thought of as the path-ordered exponential of the 
connection along the link $e_i$ $(i=1, \ldots N_1)$. 
The collection of all these configuration spaces corresponding to 
every lattice gives an {\em over-complete} description of 
$\overline{\cal A}$. It is possible to keep track of all the repetition 
by means of a projective structure. 
We say that lattice $L$ is a refinement of lattice $L^\prime$ 
($L\geq L^\prime$) if every link $e\in L^\prime$ either 
 $e=e_1$ or  $e=e_1\circ \ldots \circ e_n$ for some 
 $e_1, \ldots , e_n  \in L$. For any pair of lattices 
related by refinement $L\geq L^\prime$ there is a projection 
$p_{L^\prime\,L}:{\cal A}_{L}\rightarrow{\cal A}_{L^\prime}$ 
\be
(g_1,g_2,g_3,\ldots,g_{N_1})
\stackrel{p_{L^\prime L}}{\longrightarrow}
 (g^\prime_1=g_2g_1,g^\prime_2,\ldots, g^\prime_{N^\prime_1}),
\ee
where $e=e_1\circ e_2$, $e\in L^\prime$, $e_1,e_2 \in L$. 

The projection map and the refinement relation have two properties 
that will allow us to define $\overline{\cal A}$ as `the configuration 
space of the finest lattice'. 
First, we can check that $p_{L\,L^\prime}\circ p_{L^\prime\,
L^{\prime\prime}}=p_{L\,L^{\prime\prime}}$. 
Second, equipped with the refinement relation `$\geq$', the set of 
embedded lattices ${\cal L}$ is a partially ordered,  directed set; i.e. 
for all $L$,  $L^\prime$ and $L^{\prime\prime}$ in ${\cal L}$ we have: 
\be
L\geq L\;\;\;;\;\;\;\;L\geq L^\prime \;\;\;{\rm and}\;\;\;L^\prime\geq L
\Rightarrow L=L^\prime\;;\;\;\;\;L\geq L^\prime\;\;\;{\rm and}\;\;\;
L^\prime\geq L^{\prime\prime}\Rightarrow L\geq L^{\prime\prime}\;;
\ee
and, given any $L^\prime,L^{\prime\prime}\in {\cal L}$, there exists
$L\in {\cal L}$ such that
\be
L\geq L^\prime\;\;\;\;{\rm and}\;\;\;\;L\geq L^{\prime\prime}.
\ee

The space $\overline{\cal A}$ is the {\it projective limit} of the
projective family, defined as follows:
\be
\overline{\cal A}:=\left\{(A_L)_{L\in{\cal L}}\in
 \times_{L\in{\cal L}}{\cal A}_L\;\;:\;\;L^\prime\geq L
\Rightarrow p_{L\,L^\prime}A_{L^\prime}=A_{L}\right\}.
\ee
That is, the projective limit is contained in the Cartesian product of
{\it all} possible configuration spaces ${\cal A}_L$, subject to the
consistency conditions stated above. There is a canonical projection $p_L$
from the space $\overline{\cal A}$ to the spaces ${\cal A}_L$ given by,
\be
p_L\;\;:\;\;\overline{\cal A} \rightarrow{\cal A}_L,\;\;\;p_L((A_{L^\prime})
_{L^\prime\in {\cal L}}):=A_L.
\ee
Given any function $f_L$
defined on the space ${\cal A}_L$ one can define a function on
$\overline{\cal A}$ via the pull-back $p^*_{L}:C^0({\cal A}_L)\rightarrow
{\rm Fun}(\overline{\cal A})$. Such functions are called {\it cylindrical},
and the space of such functions is denoted by ${\rm Cyl}({\cal A})$. 
In a similar fashion, we can define
Hilbert spaces for each space ${\cal A}_L$ and `pull them back' to
the projective limit. That is, we have a projective family of
Hilbert spaces ${\cal H}_L$.

The final picture is that the kinematical Hilbert space is the space
of square integrable functions on $\overline{\cal A}$ with respect to
a measure $\mu$, that is,
\be
{\cal H}_{\rm kin}:=L^2(\overline{\cal A},d\mu).
\ee

In the case of compact gauge groups $G$, the uniform Haar measure
$\mu_0$ on each copy of the group corresponding to the links, endows
the Hilbert space ${\cal H}_{\rm kin}$ with a probability measure that is 
diffeomorphism invariant (also denoted by $\mu_0$).

Cylindrical functions define a dense subset on the Hilbert space.
It is important to point out that even though the states are associated
to the lattice, the natural labels for them are graphs $\gamma$ on $L$.
A (closed) graph $\gamma$ is just  a  collection of links of the lattice
$L$. We shall therefore denote the states by $\Psi_{\gamma}$, without
explicitly writing down the lattice $L$ where $\gamma$ lies.  Note
that in this representation the connection is `diagonal'. 
More precisely,
given a function $\Psi_\gamma(\bar{A})$ defined on a lattice $L$, for example
the trace of the holonomy $T_\eta$ along a loop $\eta$ contained
in $L$, the corresponding operator will act by multiplication:
\be
(\hat{T}_\eta\cdot\Psi_\gamma)(\bar{A}):=T_\eta(\bar{A})\Psi_\gamma(\bar{A}).
\ee

The next step involves defining the constraints on this Hilbert space
${\cal H}_{\rm kin}$; in practice we do it by defining the action of the 
constraints on ${\rm Cyl}({\cal A})$ and extend it to the whole 
${\cal H}_{\rm kin}$.  The first constraint we will 
consider is the Gauss constraint, that generates gauge
transformations. A finite gauge transformation takes the 
holonomy $g_1$ to $g(\a)g_1g(\b)^{-1}$ (where edge $e_1$ goes from vertex 
$\a$ to vertex $\b$). The quantum Gauss constraint should impose gauge
invariance on the wave functions; therefore, on a cylindrical function, 
$\Psi_\gamma(\bar{A})=\psi(g_1,\ldots,g_{N_1})$, the Gauss constraint takes
the form, for each vertex $\a$ of the lattice $L$:
\be
\sum_{e_\a}X^I_{e_\a}\cdot \Psi_\gamma=0,
\ee
where $X^I_{e_{\a i}}$ is a right or left invariant vector field on the
internal direction $I$ depending on whether the link $e_{\a i}$ is incoming
or outgoing, respectively. 
Gauge invariant functions on $L^2(\overline{\cal A},d\mu)$
can be seen as functions on the quotient space of $\overline{\cal A}$ by
the space of generalized gauge transformation $\overline{\cal G}$. Thus,
$L^2(\overline{\cal A/G},d\mu)$ is the Hilbert space of solutions to the
Gauss constraint.

It turns out that $L^2(\overline{\cal A/G},d\mu)$ is not well suited to 
define the generator of diffeomorphisms (see \cite{alm2t}); instead,  
we represent {\it finite} diffeomorphisms as unitary operators on the
Hilbert space. Let us denote by $\varphi$ a finite diffeomorphism on $\Sigma$.
We want to define $\hat{U}_\varphi:L^2(\overline{\cal A/G},d\mu)\rightarrow
L^2(\overline{\cal A/G},d\mu)$, to be a unitary operator. 
For any diffeomorphism
invariant measure $\mu$ we can define such representation, it will be
given by,
\be
\hat{U}_\varphi\cdot\Psi_\gamma(\bar{A}):=\Psi_{\varphi^{-1}(\gamma)}(\bar{A}).
\ee
That is, the induced action of $\varphi \in {\rm Diff}(\Sigma)$
is to `move' the graph $\gamma$ by the inverse diffeomorphism.
Solutions to the diffeomorphism constraint
will be those states that are left invariant by the unitary
operators $\hat{U}_\varphi$. These states will {\it not} belong to the
kinematical Hilbert space ${\cal H}_{\rm kin}$.
This is a general feature of the Dirac
quantization method for constrained systems, whenever the `gauge group'
generated by the constraint to be imposed is non-compact.

Solutions to the diffeomorphism constraint are distributions,
living on the dual space $\Phi^\prime$ to the dense sub-space
$\Phi={\rm Cyl}(\amodg)$ of gauge invariant, cylindrical functions.
An element $\bar\phi$ of $\Phi^\prime$ is said to be a solution to the
diffeomorphism constraint if
\be
\bar\phi[\hU\circ\psi]=\bar\phi[\psi]\;\;\;\forall\;\; \varphi\in
{\rm Diff}(\Sigma)\;\;\;\;{\rm and}\;\;\;\;\psi\in \Phi,
\ee
where `$[{\;\;}]$' denotes the dual action, $\phi[\psi]=\int_{\amodg} 
d\mu\,\bar{\phi}\psi$.

One can construct such distributions by `group averaging' over the 
group ${\rm Diff}(\Sigma)$. The infinite 
size of ${\rm Diff}(\Sigma)$ makes a precise definition of the 
group average procedure very subtle; here we follow the procedure 
given in \cite{alm2t}. 
An inner product for the space of solutions is given by the 
same formula that defines the group averaging; therefore, 
a summation over all the elements of ${\rm Diff}(\Sigma)$ 
would yield states with infinite 
norm. In this sense, prescribing an adequate definition for the 
averaging over the group ${\rm Diff}(\Sigma)$ involves some
`renormalization.' We shall give the details below.

An appropriate definition of the group 
averaging procedure follows from two observations. First, 
the inner product between two states induced by the cylindrical 
functions $f_\c,g_\d \in \Phi$ must be zero unless 
there is a diffeomorphism 
$\varphi_0 \in {\rm Diff}(\Sigma)$ that connects the two graphs 
$\c$ and $\d$ ($\c = \varphi_0 \d$). 
Second, the construction of generalized connections assigns group elements 
to un-parametrized edges. Therefore, two diffeomorphisms that restricted 
to a graph $\c$ are equal except for a reparametrization of the edges of 
$\c$ should be counted only once in the construction of group averaging 
of states based on graph $\c$. 
Thus, given a cylindrical function $f_\c\in \Phi$, we 
define $F_\c\in \Phi^\prime$ by 
\be 
F_\c[g_\d]:= \d_{[\c] [\d]} \sum_{[\varphi]\in {\rm GS}(\c)}\langle 
\hat{U}_{\varphi\cdot \varphi_0}\circ f_\c|g_\d \rangle \label{d}
\ee 
where $\d_{[\c] [\d]}$ is non vanishing only if there is 
a diffeomorphism $\varphi_0 \in {\rm Diff}(\Sigma)$ 
that maps $\c$ to $\d$; and 
$\varphi\in {\rm Diff}(\Sigma)$ is any element in the class of 
$[\varphi]\in {\rm GS}(\c)$. The discrete group ${\rm GS}(\c)$ is the group of 
symmetries of $\c$; i.e. elements of ${\rm GS}(\c)$ are maps between the 
edges of $\c$. The group can be constructed from subgroups of 
${\rm Diff}(\Sigma)$ as follows: 
${\rm GS}(\c)= {\rm Iso}(\c)/{\rm TA}(\c)$ where 
${\rm Iso}(\c)$ is the subgroup of 
${\rm Diff}(\Sigma)$ 
that maps $\c$ to itself (the `isotropy group'), and the elements of 
${\rm TA}(\c)$ are the ones that preserve all the edges of $\c$ separately (the
`trivial action' subgroup). 

The space of diffeomorphism invariant states constructed using formula 
(\ref{d}) inherits an 
inner product that makes it a Hilbert space ${\cal H}_{\rm diff}$. 

In  ${\cal H}_{\rm kin}$, two cylindrical functions 
$\phi(\bar{A})=\phi(g_1, \ldots , g_N)$, 
$\phi ^\prime (\bar{A})=\phi(g_1^\prime, \ldots , g_N^\prime)$ 
are {\it orthogonal} unless all the edges that define them 
are the same. In passing to
diff-invariant distributions via the group average, we are identifying
states that are associated to graphs that are diffeomorphic. 
Furthermore, we are also identifying  
the Hilbert spaces associated to lattices that are related by a 
diffeomorphism. In the case when $L=\varphi_0\circ L^\prime$, 
functions $f\in{\cal H}_L$ and $f^\prime\in {\cal H}_{L^\prime}$ satisfying 
$f=\hat{U}_{\varphi_0}\circ f^\prime$ 
will give rise to the {\it same} $F\in \Phi^\prime$. Let us denote 
by  ${\cal H}_{[L]}$ the Hilbert space of states 
associated to the diffeomorphism equivalence classes of lattices $[L]$: 
\be
{\cal H}_{[L]}:=\left\{F\in\Phi^\prime\;\;:\;\;F=\sum_{[\varphi]\in{\rm GS}(L)}
\sum_{\rm S(L)}
\hat{U}_{\varphi}\hat{U}_{\varphi_0} \circ f_L\;\;\;
 \forall\;\;\; L\in[L]\right\}
\ee 
where $\varphi_0 \in {\rm S}(L)$ is an element of ${\rm Diff}(\Sigma)$
that relates $L$ and $L^\prime \in [L]$ ($L=\varphi_0 \circ L^\prime$). 
${\rm S}(L)$ contains only one such element relating the two lattices.

Before proceeding we would like to make the following remark: 

\ni
{\em As stated above the cylindrical functions are naturally labeled by 
graphs rather than by lattices; therefore a distribution 
$F\in \Phi^\prime$ can come from the group average of cylindrical 
functions $p_L^*f, p_{L^\prime}^*f^\prime \in \Phi$ that come from 
lattices that are not diffeomorphic 
$f\in {\cal H}_L, f^\prime \in {\cal H}_{L^\prime}$, $L\not\sim L^\prime$.} 
Also, using the same line of reasoning, one can see that 
a diffeomorphism that does not map a whole lattice to itself 
can identify states that were different at the non-diffeomorphism 
invariant level. This subtlety will play an important role in the
comparison between the quantizations on the continuum and
the abstract lattice.

The Hilbert space of diff-invariant states ${\cal H}_{\rm diff}$ will
be generated by the Hilbert spaces ${\cal H}_{[L]}$. However, in the
case of finite graphs, the Hilbert space was constructed by taking
a direct sum of Hilbert spaces of equivalence classes of graphs. On the
other hand, in the
lattice construction, the Hilbert spaces of two different classes of
lattices have a non-trivial intersection 
(due to the subtlety mentioned before), so the sum is not direct.
Thus, ${\cal H}_{[L]}$ {\it do} span the Hilbert space
${\cal H}_{\rm diff}$, but we have lost the orthogonal decomposition that
was available for graphs.

We should emphasize that 
although one is working with finite lattices and the Hilbert spaces
defined on them, there is no `continuum limit' to be taken. By 
considering all the possible finite lattices we are
{\it already} in the continuum.

\subsection{Quantization on the Lattice}

Let us recall that in the ordinary lattice formulation of gauge theories,
we have {\it one} lattice on which the theory is defined. 
The continuum limit is recovered by taking the limit of the
`lattice spacing' going to zero.
When dealing with diffeomorphism invariant theories one has to
be more careful since the notion of lattice spacing does not have an
invariant meaning. One has to device more subtle ways of recovering the
continuum theory. 

Given the abstract lattice $L_{0}$, the configuration space
${\cal A}_{L_0}$ is given by
$N_1$ copies of the group $G$, corresponding to the 
matrices of parallel transport 
along the $N_1$ links of the lattice. Coordinates for the `momenta'
are naturally given by elements of the Lie algebra $g$ that label the 
right or left invariant vector fields $X^I_{e_i}$ on the $e_i$ copy
of the gauge group $G$.
The natural representation is to choose  wave functions $\Psi$ on the
 $N_1$ copies of the group: $\Psi=\psi(g_1,g_2,
\ldots,g_{N_1})$. The Hilbert space is 
${\cal H}_{L_0}:=L^2({\cal A}_{L_0},d\mu)$, where the measure 
$\mu$ is given by the Haar measure. The momenta $E$ are represented by,
\be
\hat{E}_{e_i}^I\cdot\Psi:=-i X^I_{e_i}
\cdot\Psi.\label{mop}
\ee
The imposition of the Gauss constraint follows very
closely that of the continuum, the condition of the wave functions being
\be
-i\sum_{e_\a}X^I_{e_\a}\cdot\Psi=0,
\ee
where $X_{e_{\a}}^I$ is a left or right invariant vector field depending 
on our choice of $E^I_{(\a\b)}$ or $E^I_{(\b\a)}$ to denote the
momenta associated to the link $(\a\b)$.

Mathematically, this representation is {\it identical} to the Hilbert
space ${\cal H}_L$ of the continuum for any embedding $L$ of the
lattice $L_0$ into $\Sigma$. One can prove that there is an 
isomorphism $F_{L_0}:{\cal H}_{L_0} \to {\cal H}_{[L]}$ from the Hilbert 
space constructed from the abstract lattice to the Hilbert space of 
diffeomorphism invariant states ${\cal H}_{[L]}$. 

But comparison of the two descriptions should be made between
the Hilbert space ${\cal H}_{\rm diff}$ and the (`continuum') limit
of the spaces 
$L^2({\cal A/G}_{L_0},d\mu)$ in which the abstract lattice $L_0$ fills 
all space. 
The fact that a definite procedure to take the continuum limit has not 
been completed does not prevent us from drawing some qualitative results. 
We now make the argument that we presented
in  part A precise, concluding that there is a residual 
diffeomorphism symmetry in the discretized theory: 

\begin{itemize}
\item There are states 
$p_L^* \phi _{\c _1}, p_L^* \phi _{\c _2}\in L^2(\overline{\cal A/G},d\mu)$ 
that are identified by ${\rm Diff}(\Sigma)$ and such that the corresponding 
states in ${\cal H}_{L_0}$ are different 
$F_{L_0}^{-1}(\phi _{\c _1})\not=F_{L_0}^{-1}(\phi _{\c _2})$ (see Fig. 1). 
This corresponds to the case discussed in Sec. III.B, where two 
states are related by a diffeomorphism and therefore, identified in the 
continuum theory, but the lattices on which they are defined are {\it not}
related by a diffeomorphism.
 The construction works for any pair of lattices $L_0$, 
$L$ such that $\c _1 , \c _2 \in L$, where $[L]=L_0$, and gives the same 
answer. This demonstrates that the continuum limit of the Hilbert spaces 
${\cal H}_{L_0}$ would be too big to be physically correct.

\end{itemize} 

This discussion allows us to reach our conclusion: when  working with a
theory defined on an abstract lattice, we still have to get rid of
the residual diffeomorphism symmetry. This is the second result of this
paper.

\section{discussion}

Let us briefly summarize our results. We have discussed the issue of 
diffeomorphisms on lattice formulations of theories of connections. 
The case of $2+1$ gravity is tailored for this comparison. 
It is a theory with only global degrees of freedom that are fully 
incorporated in the lattice theory even before taking the continuum limit. 
The imposition of the continuum limit is in this case almost trivial
allowing us to draw general qualitative conclusions without the 
difficulties that one encounters in $3+1$ lattice gravity \cite{renata}.
In the case of the Waelbroeck-Witten formulation, 
after imposing the curvature constraint (\ref{p=0}) the
resulting quantum theory, just as in the case of the continuum, will be
given by gauge invariant functions having support on flat connections. 
For  the Ashtekar
constraints, one expects the quantum theory to be  closely related to the
flat case. 
This expectation is based on the fact that for non-degenerate $E$'s 
(geometrodynamic sector) the two sets of constraints are equivalent. 
 In the classical theory only if both the `diffeomorphism' (\ref{h}) and 
the `Hamiltonian' (\ref{ha}) constraints are imposed, the theory describes 
flat connections and has  the correct $6g-6$ topological degrees of freedom.
This example shows that 
there {\em is} a residual diffeomorphism invariance and it has to 
be taken care of by means of more constraints. If we only consider
the `Hamiltonian constraint' (\ref{ha}), and do not impose the quantum
`translation constraints' (\ref{h}), we will end up with a theory very
far from the one describing flat connections; it will
have {\it local} degrees of freedom.

We have reviewed, in certain detail, 
the quantization of the continuum theory in order to show that,
mathematically, the quantization on the abstract lattice is equivalent to
a certain {\it subspace} of the Hilbert space of the continuum. Using this
framework, we could give examples of states that are identified in the
continuum theory, but that are not automatically identified in the 
quantization of a theory based on an abstract lattice.
Thus, this leads us to conclude that the quantum lattice theory has a residual
diffeomorphism symmetry that should be taken care of.

\section*{acknowledgments}

We would like to thank R. Loll, J. Pullin and specially A. Ashtekar
for discussions and suggestions. Both authors were supported by
Universidad Nacional Aut\'onoma de M\'exico (DGAPA, UNAM). 
This work was supported in part by  NSF grants PHY-9396246, PHY-9423950,
PHY-9514240, by the Eberly research fund  of Penn  State University 
and the Alfred P. Sloan foundation.

\section*{Appendix} 
Consider a square lattice with periodic boundary conditions, i.e. 
a grid of $N$ rows by $M$ columns on a planar representation of $T^2$. 
If $N$ and $M$ are odd, then $B(v_{ij})=0$ for every vertex $v_{ij}$ of 
the lattice implies $P(f_{ij})=0$ for every face $f_{ij}$ of the lattice. 
That follows after proving that the system of equations defined by 
$B(v_{ij})^I:=\sum_{v_{ij} \in f_{kl}} P(f_{kl})^I$ can be inverted. 
The formula for $P=P(B)$ is 
\be 
P(f_{kl})^I=\frac{1}{4} \sum_{ij} C_{kl}(ij) B(v_{ij})^I 
\label{p=p(b)}
\ee 
where $C_{kl}(ij)=\pm 1$ and the sign is given in figure (2a). 

On the other hand, if $N$ or $M$ are even, the system is not invertible and 
$B(v_{ij})^I =0$ does not imply $P(f_{ij})^I =0$. 
Now we see why invertibility 
requires an odd number of rows. In formula (\ref{p=p(b)}) we would have 
for $P(f_{11})^I$ that the coefficients $C_{11}(ij)$ 
with $i=1,2,\ldots ,M$ and $j=1,2,\ldots ,N$ are given by: 
$C_{1N}$, $C_{MN}$, $C_{M1}$, $C_{11}=-C_{M1}+4-C_{1N}-C_{MN}$, $C_{M2}$, 
$C_{12}=-C_{M2}-(4-C_{1N}-C_{MN})$, ..., 
$C_{1,N-1}=-C_{M,N-1}+(-1)^{N}(4-C_{1N}-C_{MN})$. With the consistency 
condition $C_{1N}+C_{MN}+(-1)^{N}(4-C_{1N}-C_{MN})=0$ that forces $N$ 
to be an odd number. The same argument shows that the number of columns 
should be odd. 

\hskip .2in\epsfxsize=6.2in\epsfysize=2.7in\epsfbox{corichi02.eps} 
\bigskip 

\ni
{\small
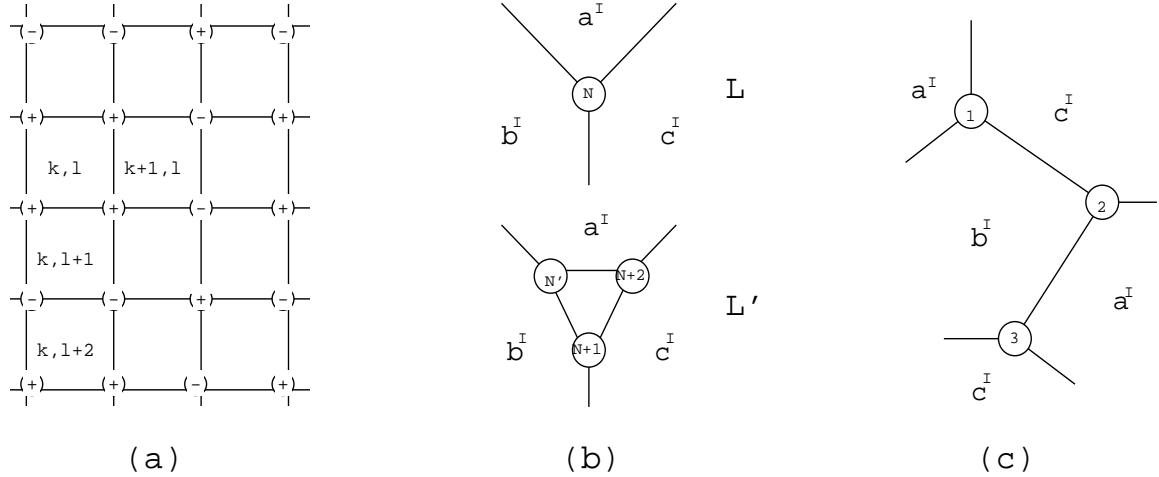
{\bf Fig. 2}a) Signs of $C_{kl}(ij)=\pm 1$. The labels of the faces are 
the ones given in the figure, and the vertices are labeled by the face 
in their immediate down-right. Every face, except for face $kl$, is 
surrounded by two plus signs and two minus signs. 
b) The lattice $L^\prime$ is generated 
from $L$ by replacing vertex $v_N$ with vertices 
$v_{N^\prime}, v_{N+1}, v_{N+2}$. 
The lattice dual to $L^\prime$ corresponds to the dual of $L$ after 
face $v_N$ has been refined by adding a vertex and three new edges in 
its interior. c)The free parameters of 
$V(L)=\{ P(f_j)^I | B(v_i)^I =0 \}$ 
are $a^I=P(f_1(v_1))^I, b^I=P(f_2(v_1))^I, c^I=P(f_3(v_1))^I=-a^I-b^I$, 
which may be subject to further conditions. 
} 

Now we study the case of a lattice with vertices of valence three. 
Again there are cases of lattices where 
$B(v_i)^I =0$ does not imply $P(f_j)^I =0$, this occurs in general 
for lattices with many symmetries. Our result is the following: 
Consider any finite three-valent lattice $L$ 
where there is a vertex $v_N$ (dual face) such that 
$B(v_N)^I=0$ but $P(f_i(v_N))^I\not= 0$ for some of the three faces 
$f_1(v_N), f_2(v_N), f_3(v_N)$ that contain vertex $v$. 
The three-valent lattice 
$L^\prime$ constructed replacing vertex $v_N$ by three vertices 
$v_{N^\prime}, v_{N+1}, v_{N+2}$ (see figure (2b)) 
is such that $B(v_i)^I =0$ implies $P(f_j)^I =0$. 

First we prove that ${\rm Dim}(V(L))\leq 2\times {\rm Dim}(su(2))=6$, 
where $V(L)=\{ P(f_j)^I | B(v_i)^I =0 \}$ is the vector space of the 
curvature vectors of all the faces $(f_j)$ of the lattice restricted 
to the condition $B(v_i)^I =0$ for every vertex $(v_i)$. 
Since $L$ is finite, we can number its vertices leaving $v_N$ at the 
end $v_1, v_2, \ldots , v_N$, and in such a way that vertex $v_R$ 
for $1<R\leq N$ is joined by an edge of $L$ to $V_S$ with 
$1\leq S<R$ (if the lattice is not connected, 
we can prove our result independently for its connected components). 
Denote the curvature vectors of the faces containing the first vertex 
by $a^I=P(f_1(v_1))^I, b^I=P(f_2(v_1))^I, c^I=P(f_3(v_1))^I$. 
$B(v_1)^I=0$ implies $a^I +b^I +c^I =0$ and $B(v_2)^I=0$ implies that 
$P(f(v_2))^I=a^I$; then we use $B(v_3)^I=0$ to see 
$P(f(v_3))^I=c^I$, etc (see figure (2c)).
In this way we see that all the free parameters 
in $V(L)$ are $a^I, b^I, c^I=-a^I-b^I$, 
which proves ${\rm Dim}(V(L))\leq 2\times {\rm Dim}(su(2))=6$. 

Now we see that ${\rm Dim}(V(L^\prime))=0$. In constructing $L^\prime$ 
from $L$ we generated only one new face (see figure(2b)). 
Therefore, the construction given above 
parameterizes the curvature vectors of all the faces of $L^\prime$ 
except for one. And the conditions $B(v_{N^\prime})^I=0$, 
$B(v_{N+1})^I=0$, $B(v_{N+2})^I=0$ demand $a^I=b^I=c^I=0$, 
which concludes our proof.

\end{document}